\documentclass[11pt,twoside]{article}

\usepackage{asp2006}
\usepackage{epsf}
\usepackage{psfig}
\usepackage{lscape}

\def\solm{M$_{\odot}\,$}

\def\solm{M$_{\odot}\,$}

\def\mass{$10^{11}$ M$_{\odot}\,$}
\def\hmass{$10^{11.5}$ M$_{\odot}\,$} 
\begin{document}
\title{Star Formation and Merging in Massive Galaxies at $z < 2$}   
\author{Christopher J. Conselice}   
\affil{University of Nottingham, England NG7 2RD}    

\begin{abstract} 
Observing massive galaxies at various redshifts is one of the most 
straightforward
and direct approaches towards understanding galaxy formation.  There
is now largely a consensus that the massive galaxy (M$_{*} >$ \mass)
population is fully formed by $z \sim 1$, based on mass and luminosity
functions.  However, we argue that the latest data can only rule out number and
mass density evolution of a factor of $> 2-3$ at $z < 1.5$.  We 
furthermore show that
the star formation history of M$_{*} >$ \mass galaxies reveals 
that 40$\pm$5\% of galaxies with M$_{*} >$ \mass at $z \sim 1$ are 
undergoing star formation that effectively doubles their stellar mass 
between $z = 0.4 - 1.4$. These massive galaxies also undergo 
0.9$^{+0.7}_{-0.5}$ 
major mergers during this same time period.
\end{abstract}

\vspace{-1cm}

\section{Introduction}

Understanding when and how massive galaxies form is one of the 
most outstanding problems in cosmology and galaxy formation.   Galaxies 
are predicted in Cold Dark Matter based models of structure formation to 
form gradually with time through the merging of smaller systems (e.g,. Cole
et al. 2001).  While there is some evidence for this 
process, in terms of galaxies (e.g., Patton et al. 2002; 
Conselice et al. 2003; Bridge et al. 2007), many details are still 
lacking. Alternatively, massive galaxies, which are mostly ellipticals 
in today's universe (e.g., Conselice 2006a), may have formed very rapidly 
in  `monolithic' collapses.

A such, massive galaxies are the dominant test-bed for galaxy models, and
understanding their evolution observationally is an important 
goal.   Most massive galaxies at $z < 1$ are red and passively
evolving, yet star formation and merging activity have been seen 
in ellipticals from $z \sim 0$ to $z \sim 1$ (Stanford et al. 2004; 
Lin et al. 2004; Teplitz et al. 2006; Conselice et al. 2007a,b) - thus it
is unclear when or how massive galaxies finally 
assembled. We must study this process at high redshift since the ages of 
stars in nearby galaxies cannot reveal their entire formation history, as
the assembly of mass is likely decoupled from the formation
of stars (e.g., Conselice 2006b; Trujillo et al. 2007). 

Observational evidence suggests that passively evolving massive galaxies 
exist at $z \sim 1$, and likely at even early times, at $z > 2$ (Daddi et al. 
2004; Saracco et al.  2005; Bundy et al. 2006; Conselice et al. 2007a).
Recent claims also exist for the establishment of the full massive galaxy
population by $z \sim 1$ (e.g., Drory et al. 2005; Bundy et al. 2006). 
However, what is not yet clear is if number densities measured in these surveys
are able to rule out evolution at $z < 1$ after considering uncertainties in 
measuring stellar masses, number densities, and cosmic variance.

 \setcounter{figure}{0}
 \begin{figure}[!h]
 \plotone{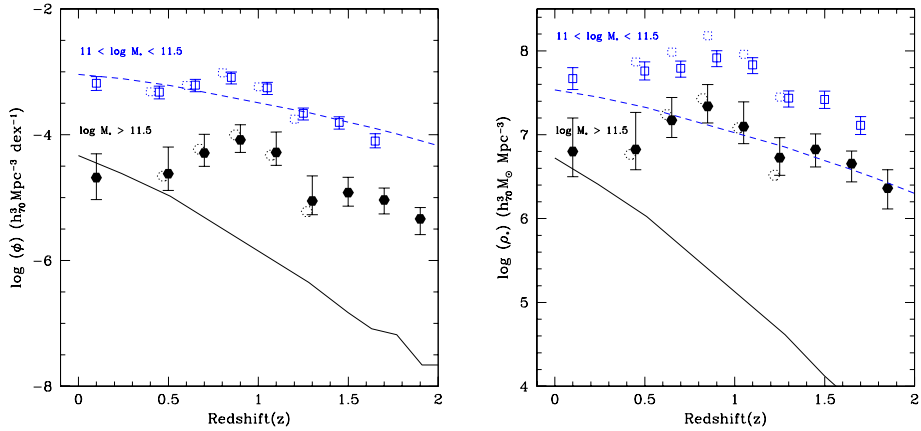}
 \caption{Left panel: the evolution in the number densities for galaxies of
various masses between $z \sim 0.4 - 1.4$.  Right panel: the
stellar mass density evolution as a function of galaxy mass at the same
redshift intervals.  The points at $z \sim 0$ are taken from Cole
et al. (2001).  The error bars listed on both the
number and mass densities reflect uncertainties from stellar mass
errors, as well as cosmic variance, and counting statistics (Conselice
et al. 2007b).  The solid and dashed lines show predictions of these
quantities from the Millennium simulation.}
\end{figure}

On the other hand, at $z > 1.5$ it appears that there are significantly 
fewer massive galaxies than at $z < 1.5$ (e.g., Drory et al. 2005).  
Observationally,  the most massive galaxies at $z > 1.5$ are undergoing 
major mergers, which are able to construct the stellar masses of massive 
galaxies rapidly (Conselice 2006b; Conselice et al. 2008a).   The situation at 
$z < 1.5$ is not as clear, with observations inconclusive on whether 
there is evolution in the massive galaxy population
at $z < 1$ (e.g., Brown et al. 2006).  
We argue here that by using a stellar mass selected sample of
galaxies the number and mass densities of massive galaxies are consistent, 
within their errors, with no evolution at $z < 1.5$, yet there is observable
evolution when examining in detail the physical processes 
occurring within these galaxies. Up to 50\% of galaxies with M$_{*} >$ \mass 
are undergoing star formation at $z \sim 1$, and roughly one major merger 
occurs within these systems at $z < 1.4$.  
We assume throughout a standard cosmology of 
H$_{0} = 70$ km s$^{-1}$ Mpc$^{-1}$, and $\Omega_{\rm m} = 
1 - \Omega_{\lambda} = 0.3$.

\section{Evolution of Massive Galaxy Number and Mass Densities}

The most basic method for calculating the evolution of 
galaxies is measuring how their number densities and integrated 
stellar mass densities change
as a function of time.   Recent work on measuring densities
suggests that within the measurement uncertainties galaxies with
stellar masses M$_{*} >$ \mass, are largely in place at $z \sim 1$ 
(Bundy et al. 2005, 2006).  Figure~1 shows the most up to date version of 
how the number and mass densities of galaxies with stellar masses  
M$_{*} > 10^{11.5}$ \solm and 
$10^{11}$~\solm~$<$~M$_{*}~<~10^{11.5}$ \solm evolve out to $z \sim 2$,
as seen in the 1.5 deg$^{2}$ Palomar Observatory Wide-Field Infrared 
Survey (POWIR; Conselice et al. 2007a; Conselice et al. 2008b) 
covering the DEEP2 fields (Davis et al. 2007) from 
Conselice et al. (2007b) using a Chabrier IMF.  The number density 
evolution of these
massive galaxies shows that statistically there is very little to no
evolution at $z < 1$ for the M$_{*} >$ \mass systems. This appears to support
the idea that nearly all massive galaxies are present by 
$z \sim 1$ (e.g., Bundy et al. 2006).

However, as can be seen within the observational errors of Figure~1, 
there is potentially evolution in number densities for M$_{*} >$ \mass
selected galaxies between $z \sim 1-1.5$.  Galaxies with M$_{*} >$ \hmass 
show an increase in number densities between $z = 1.5$ to 0.4 of a factor 
of 2.7$^{+1.8}_{-1.7}$. This is
however significant only at the  $< 2 \sigma$ level, considering all 
uncertainties (Conselice et al. 2007b). A similar result is found for the 
mass densities of galaxies with M$_{*} >$ \hmass, and it is
impossible to rule out that massive galaxies with M$_{*} >$ \hmass are
all in place at $z < 2$ (Conselice et al. 2007b).
However, as for the star-formation downsizing described in e.g.,
Bundy et al. (2006) there is also a stellar mass downsizing, such that
lower mass galaxies are formed after higher mass ones.
The number densities of systems with $10^{11}$ \solm $<$ M$_{*} < 10^{11.5}$ \solm increases
by a factor of 2.2$^{+0.57}_{-0.41}$ between $z = 1.4$ and $z = 0.4$, 
a result significant at $>$ 4~$\sigma$. Just as for the most massive
systems, this evolution occurs completely at $z > 1$ (Conselice et al. 2007b).
Taken as a whole, we calculate that the scenario whereby
the stellar mass and number densities of massive galaxies does not evolve 
between $z \sim 1.5$ to $z \sim 0.4$ can be rejected at $> 8$ $\sigma$
confidence. Through a careful analysis of number densities it does not 
appear that high mass galaxy formation, with the possible exception of 
M$_{*} >$ \hmass systems, is complete by $z \sim 1.5$. 
There could be  a factor of $2-3$ evolution in the mass and number
densities for massive galaxies at $z < 1$ due to measurement uncertainties. 
We note that the Millennium 
simulation underpredicts the densities of these massive galaxies by
a large factor (Figure~1), demonstrating how difficult it is to model 
these systems.

\section{Star Formation in Massive Galaxies at $z < 2$}

One way to understand the star formation history of massive galaxies
is to examine their position on color-magnitude diagrams. Conselice
et al. (2007b) find that 35$\pm$4\% of galaxies with M$_{*} >$ \mass
at $z \sim 1$ are blue in color and not on the red sequence, and thus must 
be undergoing unobscured star formation.     Previous studies have
examined the increase in the amount of stellar mass on the red-sequence,
finding as much as a factor of two increase 
since $z \sim 1$ (Faber et al. 2007; Brown et al. 2007). 
However, this increase is due to galaxies appearing on the red-sequence,
which were previously blue, and not due to in-situ growth on
the red-sequence itself due to e.g., dry mergers. This can be seen
through massive galaxies  gradually moving onto the red-sequence with
time, as the number of blue massive galaxies declines at $z < 1$ (e.g.,
Bundy et al. 2006; Conselice et al. 2007b).  

 \setcounter{figure}{1}
 \begin{figure}
 \plotfiddle{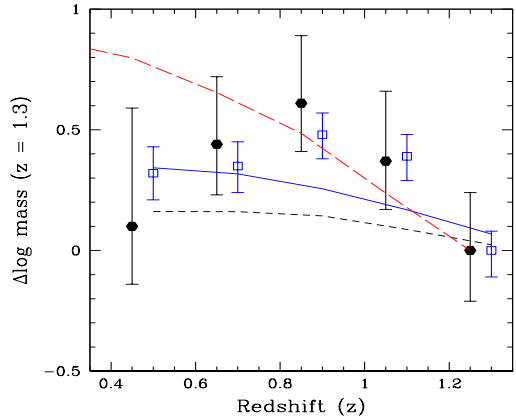}{5cm}{0}{35}{30}{-120}{-40}
 \caption{The evolution in mass density for M$_{*} >$ \mass galaxies
as a function of redshift and stellar mass.  The solid dots and open boxes
represent the evolution of galaxies with stellar masses  M$_{*} >$ \hmass
and \mass $<$ M$_{*}$ $<$ \hmass, respectively, referenced to
their stellar mass densities at $z \sim 1.3$.  The solid
blue line shows the evolution in the amount of stellar
mass added from the $z \sim 1.3$ bin as a function of redshift from
galaxies with stellar masses of \mass $<$ M$_{*}$ $<$ \hmass.
The short-dashed line shows a similar increase in stellar
mass due to star formation for the M$_{*} >$ \hmass systems.  The 
long-dashed red line shows the relative increase in the stellar
mass for the M$_{*} >$ \hmass systems due to stellar mass brought up
from the \mass $<$ M$_{*}$ $<$ \hmass bin due to star formation (Conselice
et al. 2007b).}
\end{figure}

Quantifying the ongoing rate of star formation in massive galaxy samples 
can be done in several ways, including rest-frame UV emission, emission 
line fluxes, and Spitzer MIPS 24$\mu$m data.  A surprisingly higher
fraction of massive galaxies are detected  in the mid-infrared
compared with the fraction which are blue.
Conselice et al. (2007b) find that $\sim$40\% of the M$_{*} >$ \mass 
systems at $0.4 < z < 1.4$ are detected in 24$\mu$m MIPS data after
removing those systems which are detected by Chandra.  A total of 
37$\pm5$\% of the systems at M$_{*} >$ \hmass are detected
at 24 $\mu$m, with an average star formation rate of 70 \solm yr$^{-1}$.

Conselice et al. (2007b) find, similar to previous studies
utilising IR star formation indicators, a decline with redshift in star 
formation occurring within the massive galaxy population.
For systems with M$_{*} >$ \hmass the star formation rate declines as 
$(1+z)^{6\pm2.2}$, and for systems with \mass $<$ M$_{*}$ $<$ \hmass as
$(1+z)^{4.1\pm0.64}$.  The overall decline in the
entire galaxy population's star formation history can be
parameterised as $\sim (1+z)^{3-4}$. 
It appears that while the \mass $<$ M$_{*}$ $<$ \hmass galaxies have a 
similar decline as the overall field, the highest mass galaxies
decline at a faster rate.

\section{Mergers in Massive Galaxies}

The number and mass densities of the most massive galaxies
are significantly lower at $z > 2$ than at lower redshifts.
While the star formation rate in these massive galaxies may be
low (e.g., Krick et al. 2006), mergers are likely the major
method for forming these galaxies. Examining the 
structural CAS and other parameters (Conselice et al. 2000,2002;
Conselice 2003; Ravindranath et al. 2006) for galaxies at $2 < z < 3$ reveals
that mergers produce the rapid  
growth in massive galaxies at $1.5 < z < 3$ (Conselice et al.
2003, 2005; Conselice 2006b; Conselice et al. 2008a). 

What is not yet clear is the role mergers play in massive
galaxy formation at $z < 1.5$, particularly dry mergers that 
are not easily identifiable through
structure.  One method of measuring the amount of
merging is by examining changes in the number and mass densities of
galaxies as a function of time after accounting for growth due to star 
formation (Figure~2; Conselice et al. 2007b).   Conselice et al. (2007b) 
find that 12\% of galaxies with \mass~$<$~M$_{*}~<$~\hmass merge 
between $z \sim 1.2 - 1.4$, and enter the higher mass 
bin.  At $z \sim 0.8 - 1.0$ this merger fraction drops to 8\%.  
This is consistent with the CAS merger fractions (e.g., Conselice et al. 
2003) for the same galaxies, and previous published results using 
smaller galaxy samples (e.g., Conselice et al. 2003; Lin et al. 2004;  
Bridge et al. 2007).  This corresponds to a merger rate giving
N$_{\rm m} = 0.9^{+0.7}_{-0.5}$ mergers for M$_{*} >$ \mass
systems at $0.4 < z < 1.4$, roughly doubling the mass of these
galaxies during this time.

\section{Summary}

While massive galaxies can be identified up to $z \sim 3$ using
deep NIR imaging, the evolution of these galaxies is difficult to
measure within a factor of 2-3 using simply luminosity or
mass functions. By examining the change in the mass function, and
through structural parameters and MIPS 24 $\mu$m and [OII] line
emission we show that massive galaxies are still undergoing some
evolution, with at least as much as a factor of two increase
in stellar mass at $z < 1.4$ from star formation and merging.



\end{document}